\documentclass[sigconf, nonacm, authorversion, review=false, timestamp=false, screen]{acmart}

\usepackage{stfloats}
\usepackage{graphicx}
\graphicspath{ {images} }
\usepackage{fancyhdr}

\AtBeginDocument{%
  }

\AtBeginDocument{%
    \addtolength{\footskip}{2.0\baselineskip}%
    \fancyfoot[L]{\textit{\textbf{Preprint --- do not distribute.}}}%
    \fancyfoot[C]{\thepage}
}
\begin{document}

\title{Learning the Market: Sentiment-Based Ensemble Trading Agents}

\author{Andrew Ye}
\affiliation{%
  \institution{Case Western Reserve University}
  \city{Cleveland}
  \state{Ohio}
  \country{USA}}
\email{ye@case.edu}

\author{James Xu}
\affiliation{%
  \institution{Case Western Reserve University}
  \city{Cleveland}
  \state{Ohio}
  \country{USA}}
\email{jhx2@case.edu}

\author{Vidyut Veedgav}
\affiliation{%
  \institution{Case Western Reserve University}
  \city{Cleveland}
  \state{Ohio}
  \country{USA}}
\email{vsv22@case.edu}

\author{Yi Wang}
\affiliation{%
  \institution{The Pennsylvania State University}
  \city{State College}
  \state{Pennsylvania}
  \country{USA}}
\email{ykw5273@psu.edu}

\author{Yifan Yu}
\affiliation{%
  \institution{University of Washington}
  \city{Seattle}
  \state{Washington}
  \country{USA}}
\email{yifany23@uw.edu}

\author{Daniel Yan}
\affiliation{%
  \institution{University of Southern California}
  \city{Los Angeles}
  \state{California}
  \country{USA}}
\email{dhyan@usc.edu}

\author{Ryan Chen}
\affiliation{%
  \institution{Mentor High School}
  \city{Cleveland}
  \state{Ohio}
  \country{USA}}
\email{105092@students.mentorschools.org}

\author{Vipin Chaudhary}
\affiliation{%
  \institution{Case Western Reserve University}
  \city{Cleveland}
  \state{Ohio}
  \country{USA}}
\email{vxc204@case.edu}

\author{Shuai Xu}
\affiliation{%
  \institution{Case Western Reserve University}
  \city{Cleveland}
  \state{Ohio}
  \country{USA}}
\email{sxx214@case.edu}


\begin{abstract}

    We propose and study the integration of sentiment analysis and deep reinforcement learning ensemble algorithms for stock trading by evaluating strategies capable of dynamically altering their active agent given the concurrent market environment. In particular, we design a simple-yet-effective method for extracting financial sentiment and combine this with improvements on existing trading agents, resulting in a strategy that effectively considers both \textit{qualitative} market factors and \textit{quantitative} stock data. We show that our approach results in a strategy that is profitable, robust, and risk-minimal -- outperforming the traditional ensemble strategy as well as single agent algorithms and market metrics. Our findings suggest that the conventional practice of switching and reevaluating agents in ensemble every fixed-number of months is sub-optimal, and that a dynamic sentiment-based framework greatly unlocks additional performance. Furthermore, as we have designed our algorithm with simplicity and efficiency in mind, we hypothesize that the transition of our method from historical evaluation towards real-time trading with live data to be relatively simple.
\end{abstract}


\received{2 August 2024}

\maketitle

\section{Introduction}
Continued success in the stock market - an ever-evolving and seemingly stochastic environment - demands the constant pursuit of efficient, profitable, and adaptive trading strategies. For example, the United States has seen considerable growth in its economy's value and scope during the last decade, with its Gross Domestic Product (GDP) surpassing that of other developed markets \citep{nathan2023goldman}. As these markets continue to grow, it has evidently become increasingly more difficult for traditional analysts to consider and leverage all relevant factors that influence the performance of equities such as stocks. As such, there has been consistent interest in utilizing advancements in artificial intelligence and deep learning algorithms, which perform well on high dimensions of data, to account for such shortcomings.

In particular, the exploration of using deep reinforcement learning to automate portfolio allocation has seen notable activity. These algorithms are powerful since they adapt well to dynamic decision-making problems, such as how much of a stock to buy or sell, provided they are given a sufficient amount of interaction with historical data. Fortunately, a plethora of such data exists for stock information through sources such as Yahoo Finance, Bloomberg, and more -- making the task of modeling the market a natural and achievable endeavor. Indeed, prior works have demonstrated success in implementing standalone and combinations of well-known deep reinforcement learning methods to conduct financial tasks, displaying higher returns and reduced risk compared to traditional quantitative methods \citep{ensemble2020, liu2022practical}. Within this, the technique of \textit{ensemble-learning} -- utilizing multiple instances of agents at once -- has seen notable success \citep{ensemble2020, li2022ensemble, yu2023dynamic}.

While such methods are certainly effective in evaluating decisions based on \textit{quantitative} data, we argue that another vital \textit{qualitative} component when evaluating trading strategies is developing an accurate understanding of current market sentiments. For instance, \citep{ensemble2020} found that deep reinforcement learning agents trained in one particular environment (e.g. bullish) may not necessarily perform the same when exposed to another (e.g. bearish). Thus, an effective and persisting strategy must consider both concurrent quantitative and qualitative factors when making decisions. Specifically, agents must, in addition to utilizing their learned algorithms to generate profit, recognize when the market inevitably shifts and subsequently redevelop their approach.

In this paper, we integrate sentiment analysis into ensemble-learning algorithms and build upon prior developments of automated stock allocation agents. We show that even a simple integration can lead to significant performance improvements, and demonstrate that our algorithm recognizes when market sentiments shift and accordingly adjusts its trading strategy to reflect such changes. Finally, we show that our method results in a trading strategy that is more profitable, robust, and risk-minimal compared to current state-of-the-art strategies when backtesting on historical data.
\section{Overview}
\subsection{The Market as a Deep Learning Environment}
The seminal introduction of deep learning methods to solve reinforcement learning problems within complex environments has spurred a transformative shift in algorithmic trading strategies and the world of stock trading at large \citep{mnih2013playing}. Rather than requiring a team of traditional analysts to perform the task of portfolio allocation, a successful reinforcement learning agent could theoretically devise and execute trading strategies without the need for external supervision. As such, the topic of automated trading agents has since garnered significant interest from financial firms  and researchers alike, as they carry the potential to dramatically reduce -- or even eliminate -- the cost of manual analysis.

In the last few years, numerous advancements have brought these visions closer to practical realization. \citep{jiang2017deep} introduced a financial model-free deep reinforcement learning framework for portfolio management, sparking efforts in exploring the use of Deep Deterministic Policy Gradient (DDPG) algorithms in training viable agents. Additionally, the release of open-source reinforcement learning libraries centered around quantitative finance, like FinRL (which currently sits at ~9k stars), have promoted the discovery and implementation of novel and effective trading algorithms \citep{finrl2020}.

Particularly, the practice of using an \textit{ensemble strategy} in training agents has shown both empirical and theoretical advantages \citep{ensemble2020,li2022ensemble,yu2023dynamic}. An ensemble strategy consists of simultaneously learning several deep reinforcement learning algorithms and using an evaluation metric to periodically select the best-performing algorithm every $n$ months, where $n$ is a hyper-parameter that remains fixed. These strategies often employ well-known methods like Deep Deterministic Policy Gradient (DDPG), Advantage Actor Critic (A2C), Soft Actor Critic (SAC), and Proximal Policy Optimization (PPO) as their agents.

In addition to automating portfolio allocation, deep reinforcement learning has also seen success in various other financial modeling tasks spanning from stock screening to market prediction. As an example, \citep{bai2023mercury} views stock screening as a reinforcement learning process, and determines the value of each stock in the market and its relationship with other stocks using hyper-graph attention.

\subsection{Language-Based Trading Agents}
The integration of sentiment and natural language processing to the stock market has similarly undergone considerable development, growing tangentially with the rise of deep learning methods. Efforts in this field have primarily focused on using information gathered from a variety of public sources (forums, newspapers, etc.) to predict the future behavior of particular stocks and the market at large. For instance, \citep{mittal2012twitter} successfully used Twitter user data to analyze public sentiment. By combining predicted future sentiment with observed values of the Dow Jones Industrial Average (DJIA), their system was able to obtain around a 75\% accuracy in predicting market movements. Since then, more advanced analysis methods have been able to reach up to 80\% accuracy \citep{kalyani2016stock, xiao2023prediction}.

It is also worth noting that amassing public language data to conduct sentiment analysis poses notable risks and limitations. For instance, mainstream data sources (social media, news articles, etc.) consistently include a combination of both relevant and irrelevant data, which affect the accuracy of the systems. As such, for language-based trading agents to be implemented at scale, attention to risk-minimization should tangentially increase compared to traditional methods. In addition, it is worthwhile to note that the use of these agents introduces a greater level of susceptibility to market manipulation, as their actions are largely dictated by public, stochastic, and freely manipulable sources. Therefore, there may be a heightened need to elevate regulatory and legal standards to preserve the safe and ethical usage of these models.
\subsection{Our Motivation}
With the current successes and advancements made in deep reinforcement learning to train effective trading agents and sentiment analysis to accurately predict market movement, we hypothesize that a natural extension of the two methodologies involves the integration of fields and propose a dynamic ensemble-based trading agent that adapts based on market sentiments. 

Current ensemble methods rely on a fixed time period to re-evaluate and select their currently chosen algorithm \citep{ensemble2020, li2022ensemble, yu2023dynamic}. However, we argue that doing so may greatly inhibit the overall performance of the overall strategy. Consider, for example, an ensemble strategy currently trading via a DDPG agent. If, at the end of the time period, the market is relatively unchanged and the agent is performing well, it would make little sense to re-evaluate and switch the chosen algorithm. Conversely, if, at some point during the agents trading period, the market environment greatly changes, then we cannot expect the same agent to perform successfully. In this scenario, it would benefit to immediately re-evaluate and select a new algorithm that performs better rather than waiting until the end of the period to do so.

Detecting market sentiment in itself also poses a challenge. Current methods mainly rely on large-scale data pre-processing and advanced machine learning methods to determine the relationship between current news sentiment and stock prices, which can be an expensive and time-consuming process. While such methods have displayed impressive accuracy, the computationally expensive means in which they are acquired may inhibit effectively trading in real-time. In practice, there likely exists an \textit{accuracy-efficiency trade-off} in which maintaining a constant ``perfect'' picture of market sentiments is impractical.

Our motivation is thus two-fold. On one end, we aim to expand upon the observation that specific reinforcement learning algorithms adhere to certain environments and develop an improved strategy that dynamically adjusts to such environmental changes rather than waiting an imposed fixed period \citep{ensemble2020}, as well as accounts for the increased risk of utilizing sentiment calculations. On the other, given the relative complexity of current sentiment analysis algorithms, we opt to devise a simple yet effective procedure to efficiently extract and evaluate sentiment to facilitate the detection of changing environments.
\section{Methodology}
\subsection{Deep Reinforcement Learning}
\begin{figure}
    \centering
    \includegraphics[scale=0.55]{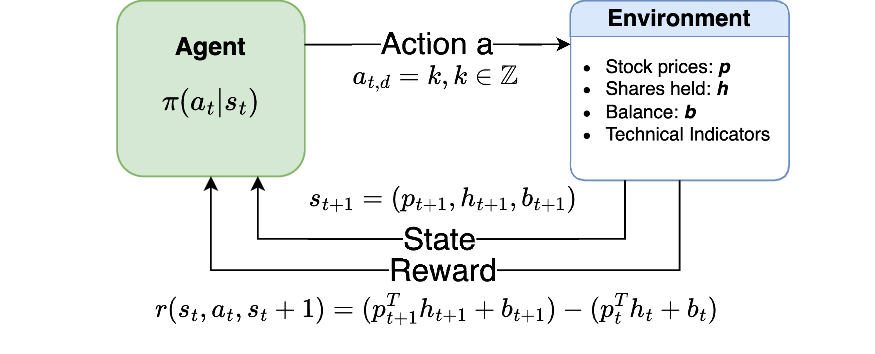}
    \caption{Stock trading as a reinforcement learning problem.}
\end{figure}

We follow the practice of 
\citep{ensemble2020} by modeling the task of stock trading as a Markov Decision Process and subsequently train three deep reinforcement learning agents within it:

\begin{enumerate}
    \item State $s=[p,h,b]$: A set including the prices of each stock $p\in \mathbb{R}^D$, the amount of each stock held in the portfolio $h\in \mathbb{Z}^D$, and the remaining balance in the account $b\in \mathbb{R}$, where $D$ denotes the total number of unique stocks considered.
    \item Action $a$: a set of actions on all stocks $a\in \mathbb{R}^D$, which includes selling $k$ shares of stock $d$ ($a_d = -k$), buying $k$ shares of stock $d$ ($a_d = k$), and holding stock $d$ ($a_d = 0$). 
    \item Reward function $r(s, a,s')$: The change in portfolio value when action $a$ is taken at state $s$ to arrive at the new state $s'$, where the total portfolio value is calculated as the sum of equities in all held stocks $p^Th$ and the remaining balance.
    \item Policy $\pi(a|s)$: A probability distribution across $a$ at state $s$ that approximates an optimal trading strategy.
    \item The action-value function $Q_\pi(s, a)$: The expected reward achieved if action $a$ is taken at state $s$ following the policy $\pi(s|a)$.
\end{enumerate}
In this environment, an agent will aim to maximize their expected cumulative change in portfolio value following several market constraints and assumptions (liquidity, non-negative balance, and transaction costs). \\
\indent We then simultaneously train three deep reinforcement learning agents in this environment, each corresponding to a different algorithm: Deep Deterministic Policy Gradient, Proximal Policy Optimization, and Advantage Actor Critic. To train and validate these algorithms, and to model the market according to the definitions above, we use the FinRL library -- an open-source framework for integrating financial reinforcement learning strategies \citep{finrl2020}.

\subsubsection{Deep Deterministic Policy Gradient} 
Deep Deterministic Policy Gradient (DDPG) is an off-policy deep reinforcement learning algorithm that concurrently learns both a Q-function and a policy gradient. It is a deep-learning extension of the Deterministic Policy Gradient algorithm (DPG) introduced by \citep{silver2014dpg}. Like its predecessor, DDPG utilizes actor and critic neural networks $\mu(s | \theta^\mu)$ and $Q(s, a | \theta^Q)$, where $\mu(s | \theta^\mu)$ is an actor-network parameterized by $\theta^\mu$ that learns the optimal action to take given state $s$ and $Q(s, a | \theta^Q)$ is a critic network parameterized by $\theta^Q$ that returns the estimated Q-value given action $a$ in state $s$. 

Like other deep Q-learning algorithms, DDPG learns a Q-function $Q(s, a | \phi)$ by minimizing the mean-squared Bellman error:
\begin{center}
    $L(D) = \mathbb{E}_{t\in D}[(Q(s_t,a_t|\phi)-y(t))^2]$,
    where \\
    $y(t)=r(s_t,a_t) + \gamma(1-d_t)max_{a'}Q(s'_t,a'|\phi)$
\end{center}
Here, $\phi$ describes the set of parameters in $Q$, and $t = (s_t, a_t, r_t,s'_t, d_t)$ describes an observation of a set of transitions in a replay buffer $D$, where $d_t$ indicates whether or not $s'_t$ is a terminal state.

Rather than solely using the original networks themselves, DDPG creates copies of the actor and critic networks, $\mu'(s|\theta^{\mu'})$ and $Q'(s, a|\theta^{Q'})$, whose parameters are slowly updated by copying over a portion of the weights in their respective original networks \citep{lillicrap2019continuous}. These copies are then used to calculate the target value $y(t)$. Thus, Q-learning in DDPG is performed by minimizing:
\begin{center}
    $L(D) = E_{t\in D}[(Q(s_t,a_t\phi)-y'(t))^2]$, \\
    where
    $y'(t)=r(s_t,a_t) + \gamma(1-d_t)max_{a'}Q'(s',\mu'(s'|\theta^{\mu'})|\theta^{Q'})$
\end{center}

Since $\mu(s|\theta^\mu)$ learns to output an action that maximizes $Q(s, a|\phi)$, traditional gradient ascent can be performed w.r.t. $\theta^\mu$ to solve
\begin{center}
    $max_{\theta^\mu} E_{s \in D}[Q(s, \mu(s|\theta^\mu)|\phi)]$.
\end{center}

\subsubsection{Proximal Policy Optimization}
Proximal Policy Optimization (PPO) is a family of policy gradient methods that aims to optimize a surrogate objective function using stochastic gradient descent \citep{schulman2017proximal}. The surrogate objective function ensures that gradient updates are not too large and that at each iteration the new learned policy is not significantly different from the previous one. The algorithm improves upon Trust Region Policy Optimization \citep{schulman2017trust} by removing the need to incorporate Kullback-Leibler divergence into its deviation penalty and allowing updates that are compatible with stochastic gradient descent \citep{schulman2017proximal}. Its objective function is defined as the maximization of:
\begin{center}
    $L^{PPO}=\mathbb{E}[min(R_t(\theta)A_t,clip(R_t(\theta), 1-\epsilon, 1+\epsilon)A_t)]$,
\end{center}
where $R_t(\theta)$ is the ratio of the probability distributions under the new and old policies:
\begin{center}
    $r_t(\theta)=\frac{\pi_\theta(a_t|s_t)}{\pi_{\theta_{old}}(a_t|s_t)}$,
\end{center}
$A_t$ is the estimated advantage at timestep $t$, and $\epsilon$ is a hyper-parameter usually between 0.1 and 0.2 \citep{schulman2017proximal}.

The integral observation of PPO is that it clips the relative amount of change in the parameters of a policy based on the estimated advantages. If, at a particular timestep, a policy's actions are worse than its prior policy, then $A_t$ is negative, and the surrogate objective function $L^{PPO}$ can at most be $(1-\epsilon)A_t$. When actions under the new policy are deemed better than the old policy, then $A_t$ is positive, and the surrogate function ensures that $L^{PPO}$ can at most be $(1+\epsilon)A_t$, ensuring that policy changes at each iteration remain conservative \citep{schulman2017proximal}. In practice, PPO is stable, fast, and simple to implement and tune.

\subsubsection{Advantage Actor-Critic}
Advantage actor-critic (A2C) is the synchronous variant of asynchronous advantage actor-critic (A3C) \citep{mnih2016asynchronous}. A2C is an actor-critic algorithm and utilizes an advantage function $A(a_t, s_t)=Q(a_t, s_t)-V(s_t)$ to reduce the variance of policy gradients. The advantage function can be reduced through the Bellman recurrence to rely only on the approximation of $V$: $A(a_t, s_t)=r(s_t, a_t) + \gamma V(s'_t)-V(s_t)$ \citep{mnih2016asynchronous}. In doing so, actions are evaluated not only on how good its raw value is but also how much better it is compared to the state baseline $V_(s_t)$. 

The gradient of the objective function of A2C is:

\begin{center}
    $\nabla L^{A2C} = E[\sum_{t=1}^T\nabla_\theta log_{\pi_\theta}(a_t|s_t)A(s_t,a_t)]$
\end{center}

Similar to A3C, A2C disperses copies of the same agent to compute its gradients concerning different data samples \citep{mnih2016asynchronous}. Each agent interacts independently with the same environment, with the key difference between A3C and A2C being that the latter is a single-worker variant of the former: A2C runs interactions with a single copy of itself and waits for the computation of all gradients in a run before averaging and updating the global parameters. The practice of synchronizing gradients is cost-effective and efficient, and has empirically shown to produce results comparable to A3C.

\subsection{Capturing Market Sentiment}

\begin{table}
    \centering
    \begin{tabular}{ |c|c| } 
     \hline
      \textbf{Period} & \textbf{Sentiment}\\ 
     \hline
     01/01/2010-03/04/2010 & -4.33 \\ 
     \hline
     03/05/2010-05/06/2010 & 0.37 \\ 
     \hline
     05/07/2010-07/08/2010 & 1.56 \\ 
     \hline
     07/09/2010-09/09/2010 & -2.20 \\ 
     \hline
     09/10/2010-11/12/2010 & 4.13 \\ 
     \hline
    \end{tabular}
    \caption{An example of captured sentiment for consecutive periods.}
\end{table}
To capture market sentiment, we write a script that gathers daily headlines from leading financial news sources (Wall Street Journal, Bloomberg, etc.) and extracts their sentiments using the AFINN-en-165 sentiment lexicon. AFINN is a family of mappings of English terms and an integer rating \textit{valence} between -5 and 5, where higher numbers indicate a more positive sentiment \citep{NielsenF2011New}. In particular, AFINN-en-165 contains 3382 entries of words and phrases and is considered a general improvement over its predecessor, AFINN-111. We include a full justification for this choice of lexicon at the end of the paper, as well as explore other choices of dictionary (\textbf{Appendix A}).

Our choice of designating news headlines as a source of sentiment stems from the general ideology that headlines must be simultaneously succinct, informative, and relevant for the sake of viewership and reputation. This makes them ideal candidates for extracting a general representation of a given subject. We focus our attention specifically on finance-related articles, since they respectively contain headlines whose sentiments coalesce into a near-accurate portrayal of the contemporary market. 

The total sentiment of a headline can thus be expressed as the average of the scores of its comprised words, and the sentiment for a given period is the average sentiment for headlines published during that period:

\begin{center}
    $Sentiment(p) = \frac{1}{|p|}\sum_{i\in p}\frac{1}{|T_i|}\sum_{k=0}^{|T_i|}score(T_i[k])$,
\end{center}

where $T_i[k]$ describes the $k$th word of the $i$th headline, $|T_i|$ describes the length (in words) of headline $i$, $p$ denotes the period, and $|p|$ denotes the number of articles published in $p$.

In contrast with other sentiment-based prediction processes, our method is simple, efficient, and easy to implement. By taking advantage of pre-existing lexicons rather than learning scores from scratch, we are able to instantaneously retrieve and calculate the average news sentiment over a period of time.

\subsection{Dynamic Learning Adaptation}
\subsubsection{Choosing an Agent}
We modify the method of \citep{ensemble2020} and determine the best-performing algorithm to be that which results in the highest relative Sharpe-Sortino ratio after validation.

The Sharpe ratio measures a portfolio's return compared to its risk, or standard deviation. It is defined as:
\begin{center}
    $Sharpe = \frac{R_p - R_f}{\sigma_p}$
\end{center}
Where $R_p$ is the portfolio return, $R_f$ is the risk-free rate, and $\sigma_p$ is the standard deviation of the portfolio. 
The Sortino ratio  is similar to the Sharpe ratio but only penalizes downside volatility. It is defined as:
\begin{center}
    $Sortino = \frac{R_p-R_f}{\sigma_d}$
\end{center}
Where $\sigma_d$ is the standard deviation of the portfolio's negative returns or downside. Thus, our final scoring metric is:
\begin{center}
    $\chi_i = \alpha*Sharpe + (1-\alpha)*Sortino$
\end{center}
Where $\alpha$ is a hyper-parameter selected before training, and $\chi_i$ denotes the post-validation performance of agent $i$. Our choice of scoring metric ensures that our agent learns to further minimize risk given its newly increased flexibility, with these improvements being supported in our experimental results (\textbf{\S4}) and ablation studies (\textbf{Appendix B}).\par

\subsubsection{Switching Agents}
Lastly, our algorithm will only switch trading agents when it has detected a change in period-to-period sentiment above a certain predetermined threshold $\beta$. The intuition behind this methodology is that because agents trained on different reinforcement learning frameworks perform differently across environments, an optimal trading strategy will derive from the practice of detecting events monumental enough to affect markets and cause a dramatic shift in public sentiment, and subsequently re-selecting the best-performing agent based on the most recent data.
\begin{figure}[t]
    \centering
    \includegraphics[scale=0.6]{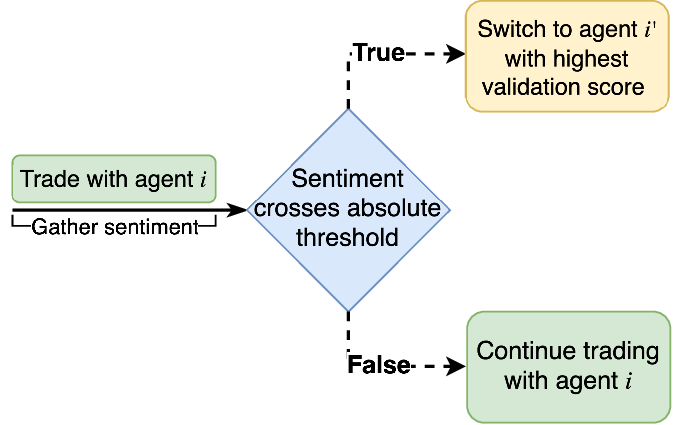}
    \caption{Overview of proposed algorithm.}
    \label{fig:enter-label}
\end{figure}

Figure 2. describes the general process in which we select agents. The first selected agent will be that which results in the highest initial validation score, $ Agent = argmax_{i\in A}(\chi_i)$. Following this, we validate agents in parallel and repeatedly gather daily news sentiments from a pre-selected database of financial sources. If the sentiment score for a given period exceeds a predetermined absolute threshold, then we switch to using the agent which currently holds the highest validation score for the given period. If the sentiment score does not exceed the threshold, then we deduce that market conditions have likely not drastically changed, and simply continue using the agent currently in selection since it has been shown to perform well in this environment.

\subsubsection{Ablation studies} Due to the various components of our algorithm (validation metrics, dynamic agent switching), it is natural to ensure that the inclusion of each part correctly achieves its intended purpose. For this reason, we conduct ablation studies (\textbf{Appendix B}) that demonstrate the contribution of each component, and find that the majority of our improvement in trading ability can be attributed to dynamic agent switching as expected, with the improved validation metric accordingly reducing risk.
\section{Evaluating Our Agent}

\begin{figure*}[t]
    \centering
    \hspace*{-0.8cm} 
    \includegraphics[scale=0.4]{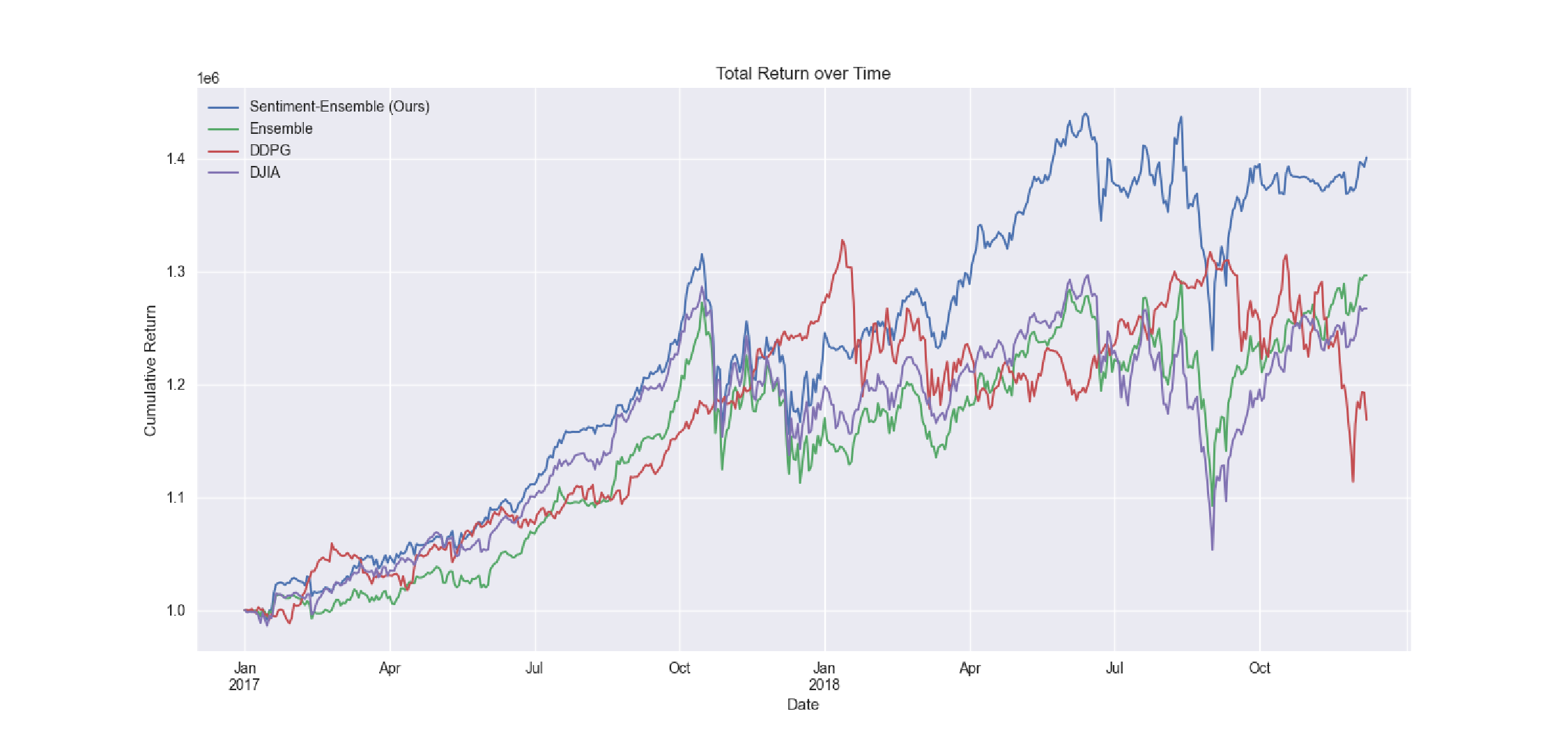}
    \caption{Performance of our algorithm (blue) against the ensemble strategy (green), a DDPG agent (red) and the Dow Jones Industrial Average (purple).}
    \label{fig:enter-label}
\end{figure*}

\begin{table*}[t]
    \centering
    \begin{tabular}{ |c|c|c|c|c| } 
     \hline
      & \textbf{Sentiment-Ensemble (Ours)} & \textbf{Ensemble} & \textbf{DDPG} & \textbf{DJIA} \\ 
     \hline
     \textbf{Cumulative Return} & \textbf{40.10\%} & 29.65\% & 16.87\%  & 17.43\% \\ 
     \hline
      \textbf{Annual Return} & \textbf{18.36\%} & 13.86\% & 8.11\% & 8.38\%\\ 
     \hline
     \textbf{Maximum Drawdown} & \textbf{-14.57\%} & -15.43\% & -16.11\%  & -18.77\% \\ 
     \hline
     \textbf{Annual Volatility} & 13.50\% & 14.49\% & \textbf{13.26\%} & 13.51\% \\ 
     \hline
     \textbf{Sharpe Ratio} & \textbf{1.32} & 0.97 & 0.66 & 0.66 \\ 
     \hline
     \textbf{Sortino Ratio} & \textbf{1.87} & 1.34 & 0.88 & 0.90 \\ 
     \hline
    \end{tabular}
    \caption{Performance metrics for Sentiment-Ensemble (our method), the conventional ensemble strategy \citep{ensemble2020}, a DDPG agent \citep{liu2022practical}, and the Dow-Jones as a benchmark. All deep-learned agents were trained on stock data from 1/1/2010 to 12/31/2016 and then evaluated from 01/01/2017 to 01/01/2019. Comprehensive results can be found in Appendix C.}
\end{table*}
We evaluate our strategy by training and testing on historical stock data - a practice commonly known as \textit{backtesting}. For comparison, we also evaluate the performance of the standard ensemble algorithm, a single DDPG agent, and the Dow Jones Industrial Average (DJIA) \citep{ensemble2020}. All deep-learning based algorithms (ours, ensemble, DDPG) are trained and tested on the same time period and data.

Specifically, our agents are trained for seven years on stock data (1/1/2010 through 12/31/2016) and then evaluated over a two-year period on out-of-sample data (01/01/2017 through 01/01/2019). For sentiment analysis, we used business articles from the Wall Street Journal Archive, with 15 articles gathered daily to generate an approximation of daily financial sentiment (although we note that our method is easily extensible to include a wider variety of financial sources). For hyper-parameters, an $\alpha$ of 0.25 was selected for measuring validation performance ($25\% $ Sharpe, $75\%$ Sortino), a $\beta$ of 15 was determined as a sentiment score threshold for switching agents, and a period was defined to be two months (62 days). These values were empirically observed to result in the most optimal trading strategy.

Table 2. shows the results of each strategy's performance during the evaluation period, and their performance relative to that of others is visualized in Figure 3. Our Sentiment-Ensemble strategy significantly outperforms both the Dow Jones and the ensemble strategy, resulting in a cumulative return of over 40\% over the two-year period and an annualized return of 18.36\%. Additionally, we find that our method results in the highest Sharpe and Sortino Ratio and the lowest maximum drawdown by a notable margin, suggesting that it learns to better manage the risk-over-return trade-off. We also note that an individual deep reinforcement strategy (DDPG) falls behind all methods, including the DJIA. We hypothesize our results indicate that traditional $n$-period agent switching is not an effective way to allocate agents in an ensemble strategy, and that a more optimal use of the method involves the introduction of utilizing market sentiment.

To confirm our hypothesis, we compare the performance of Sentiment-Ensemble (our method) and ensemble against the individual algorithms that both strategies comprise: DDPG, PPO, and A2C (Table 3). In addition, we evaluate an algorithm separate from the ones learned in ensemble, Twin Delayed DDPG (TD3) \citep{fujimoto2018addressing}. We find that although ensemble performs better than its single-agent counterparts, its performance significantly increases when paired with dynamic sentiment analysis, with the proposed method outperforming every other strategy despite leveraging the exact same agents as ensemble. With a closer examination of Figure 4, we notice that Sentiment-Ensemble appears to better merge the behavior of its component algorithms -- the goal of ensemble-learning -- before subsequently outperforming them.

\begin{figure*}
    \hspace*{-1.25cm} 
    \includegraphics[scale=0.4]{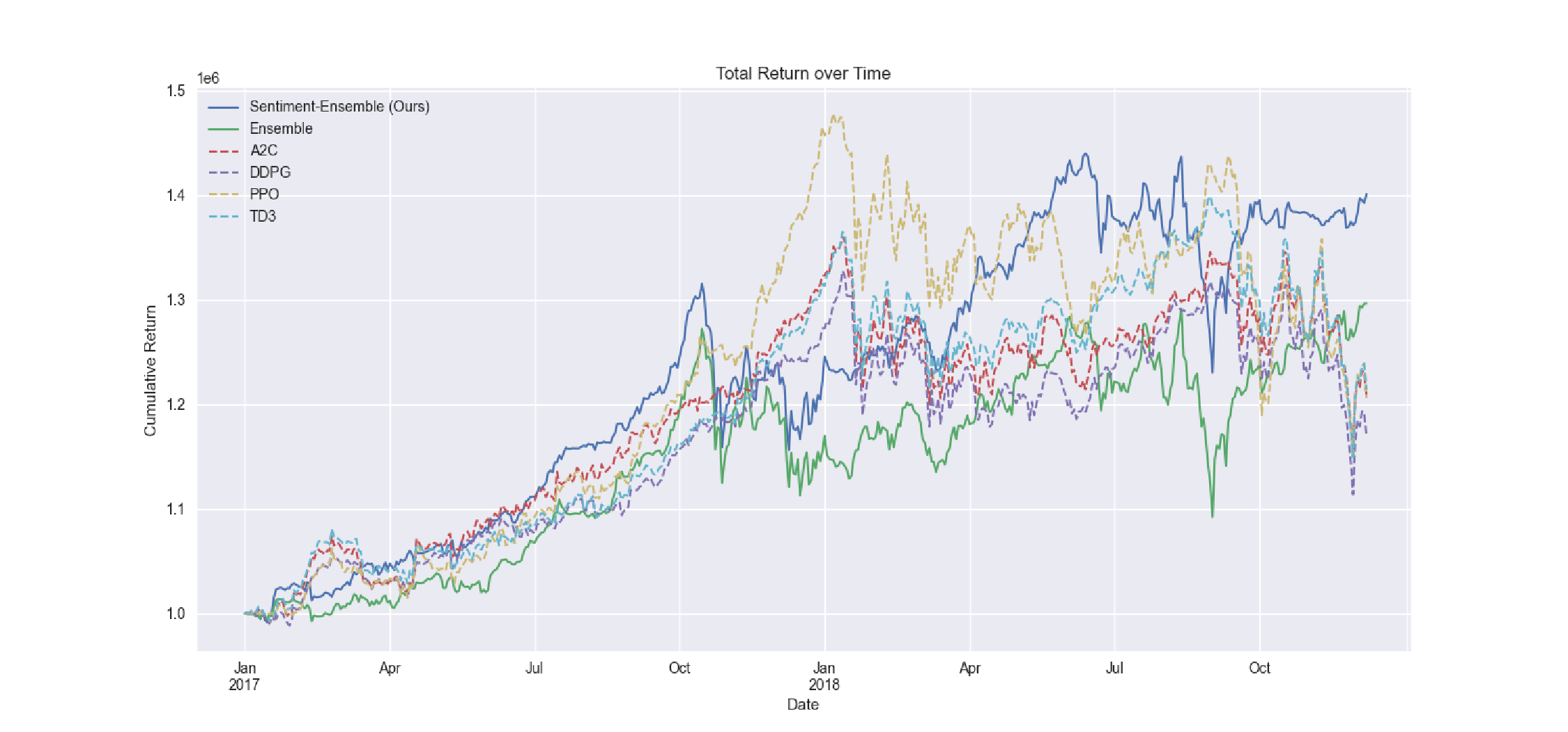} 
    \caption{Return over time of Sentiment-Ensemble (blue) and ensemble (green) against A2C (dashed-red), DDPG (dashed-purple), PPO (dashed-yellow), and TD3 (dashed-blue).}
    \label{fig:enter-label}
\end{figure*}

\begin{table*}[H]
    \centering
    \begin{tabular}{ |c|c|c|c|c|c|c| } 
     \hline
      & \textbf{S-E} & \textbf{Ensemble} & \textbf{DDPG} & \textbf{A2C} & \textbf{PPO} & \textbf{TD3} \\ 
     \hline
     \textbf{Cumulative Return} & \textbf{40.10\%} & 29.65\% & 16.87\%  & 20.81\% &  20.59\% & 21.43\% \\ 
     \hline
      \textbf{Annual Return} & \textbf{18.36\%} & 13.86\% & 8.11\% & 9.91\% & 9.81\% & 10.20\% \\ 
     \hline
     \textbf{Maximum Drawdown} & \textbf{-14.57\%} & -15.43\% & -16.12\%  & -15.05\% & -22.35\% & -17.84\% \\ 
     \hline
     \textbf{Annual Volatility} & 13.50\% & 14.49\% & \textbf{13.26\%} & 14.10\% & 17.23\% & 14.08\% \\ 
     \hline
     \textbf{Sharpe Ratio} & \textbf{1.32} & 0.97 & 0.66 & 0.74 & 0.63 & 0.76 \\ 
     \hline
     \textbf{Sortino Ratio} & \textbf{1.87} & 1.34 & 0.88 & 1.03 & 0.87 & 1.05 \\ 
     \hline
    \end{tabular}
    \bigskip
    \caption{Performance metrics for Sentiment-Ensemble and ensemble \cite{ensemble2020} against DDPG \cite{liu2022practical}, A2C \cite{mnih2016asynchronous}, PPO \cite{schulman2017proximal}, and TD3 \cite{fujimoto2018addressing}. All deep-learned agents were trained on the same stock data from 1/1/2010 to 12/31/2016 and then evaluated from 01/01/2017 to 01/01/2019. Comprehensive results can be found in Appendix C.}.
\end{table*}

\section{Conclusion}

To the best of our knowledge, we present the first successful integration of sentiment analysis into deep-reinforcement learning ensemble strategies designed for live stock trading. To achieve this, we introduce a simple-yet-effective methodology of extracting market sentiment, and incorporate this into a strategy that combines qualitative market properties with agents trained on quantitative data. Our method results in a strategy that outperforms the current state-of-the-art ensemble approach, showing that the conventional method of switching agents every $n$ months, where $n$ is fixed, is not as effective as dynamically switching agents based on environmental changes. In addition, our algorithm for extracting market sentiment is simple and efficient, and we encourage the transition of our method into real-time trading on live data, as it should be a relatively simple to implement.

Some possible future directions of interest involve exploring more sophisticated sentiment-detection strategies that employ a similar level of efficiency, such as performing inference with external large language models (LLMs) that may potentially capture market dynamics better. Additionally, the threshold for sentiment ($\beta$) could be more effectively learned, rather than set, by an external ``management'' network, which learns to manage and disperse its respective agents.

\bibliographystyle{ACM-Reference-Format}
\bibliography{bibfile}

\appendix
\section{Choice of Lexicon}
The selection of AFINN-en-165 as our lexicon-of-choice is based on the following:

\begin{enumerate}
    \item The lexicon utilizes a simple numeric scoring metric from -5 to 5, making it easier for algorithms to interpret and implement compared to other dictionaries which often score categorically \cite{loughran-mcdonald}.
    
    \item The lexicon is proven to perform well on informal text. This allows for greater accuracy, given that our dataset is comprised of news headlines which are susceptible to contain colloquialisms and abbreviations. 
    
    \item The lexicon only contains 3,382 entries of words and phrases, which is significantly lower than its alternatives which can contain, on average, over 7,000 entries \cite{Hutto_Gilbert_2014}. This can potentially improve the sentiment analysis model in the context of stock trading by introducing bias, reducing variance, and thus reducing the likelihood of overfitting, which leads to better general performance across various market conditions.
\end{enumerate}

Some other interesting lexicons we mention for further exploration include the Valence Aware Dictionary and Sentiment Reasoner (VADER), which is specifically tuned for social media  \cite{Hutto_Gilbert_2014}, and the Loughran-McDonald sentiment lexicon, which specializes in classifying financial documents \cite{loughran-mcdonald}.

\section{Ablation Study of Method}
We demonstrate here that the majority of the performance of our algorithm is attributed to the practice of dynamically assigning agents based on current market sentiment, and not on specific evaluation environments or the inclusion of the Sortino ratio by conducting ablation studies on its components. To showcase this, we evaluate the effectiveness of the Sentiment-Ensemble strategy without a) the inclusion of the Sortino ratio and b) the dynamic assignment of agents. Results can be found on the page below in Table B1 (note that Sentiment-Ensemble with \textit{both} components is our full proposed algorithm, and Sentiment-Ensemble \textit{without} either component is simply the traditional ensemble strategy).

Our results indicate that both components accordingly address their purpose. Both strategies outperform the ensemble baseline with the method trained purely on learning market sentiment and switching agents having higher overall returns at the expense of increased risk, and the method utilizing additional financial metrics having lower returns but being an overall safer strategy.
\setcounter{table}{0}
\renewcommand{\thetable}{B\arabic{table}}
\begin{table*}
    \centering
    \begin{tabular}{|c|c|c|}
        \hline
        & \textbf{S-E (w.o. Sortino)} & \textbf{S-E (w.o. DS)} \\
        \hline
         \textbf{Cumulative Return} & \textbf{39.47\%} & 34.53\% \\
         \hline
         \textbf{Annual Return} & \textbf{18.10\%} & 15.99\% \\
         \hline
         \textbf{Maximum Drawdown} & -15.49\% & \textbf{-12.26\%} \\
         \hline
         \textbf{Annual Volatility} & 14.71\% & \textbf{12.55\%} \\
         \hline
         \textbf{Sharpe Ratio} & 1.21 & \textbf{1.25} \\
         \hline
         \textbf{Sortino Ratio} & 1.71 & \textbf{1.74} \\
         \hline
    \end{tabular}
    \caption{Ablation study of the Sentiment-Ensemble (S-E) method without the Sortino ratio and dynamic switching (DS).}
    \label{tab:my_label}
\end{table*}

\section{Full results}
The full performance of statistical metrics used to evaluate strategies can be found on the page below (Table C1). We hope these results further enforce the effectiveness of our method, and highlight that the Sentiment-Ensemble strategy attains the most desirable values across all but one metric (annual volatility).
\setcounter{table}{0}
\renewcommand{\thetable}{C\arabic{table}}
\begin{table*}
    \centering
    \begin{tabular}{ |c|c|c|c|c|c|c|c| } 
     \hline
      & \textbf{S-E} & \textbf{Ensemble} & \textbf{DDPG} & \textbf{A2C} & \textbf{PPO} & \textbf{TD3} & \textbf{DJIA}\\ 
     \hline
     \textbf{Cumulative Return} & \textbf{40.10\%} & 29.65\% & 16.87\%  & 20.81\% &  20.59\% & 21.43\% & 17.43\& \\ 
     \hline
      \textbf{Annual Return} & \textbf{18.36\%} & 13.86\% & 8.11\% & 9.91\% & 9.81\% & 10.20\% & 8.38\% \\ 
     \hline
     \textbf{Maximum Drawdown} & \textbf{-14.57\%} & -15.43\% & -16.12\%  & -15.05\% & -22.35\% & -17.84\% & -18.77\% \\ 
     \hline
     \textbf{Annual Volatility} & 13.50\% & 14.49\% & \textbf{13.26\%} & 14.10\% & 17.23\% & 14.08\% & 13.51\% \\ 
     \hline
     \textbf{Sharpe Ratio} & \textbf{1.32} & 0.97 & 0.66 & 0.74 & 0.63 & 0.76 & 0.66 \\ 
     \hline
     \textbf{Sortino Ratio} & \textbf{1.87} & 1.34 & 0.88 & 1.03 & 0.87 & 1.05 & 0.90 \\ 
     \hline
     \textbf{Calmar Ratio} & \textbf{1.26} & 0.90 & 0.50 & 0.66 & 0.44 & 0.57 & 0.45 \\
     \hline
     \textbf{Omega Ratio} & \textbf{1.30} & 1.21 & 1.13 & 1.15 & 1.13 & 1.16 & 1.14 \\
     \hline
     \textbf{Tail Ratio} & \textbf{1.03} & 0.86 & 0.85 & 0.87 & 1.00 & 0.92 & 0.83 \\
     \hline
     \textbf{Stability} & \textbf{0.90} & 0.82 & 0.77 & 0.77 & 0.65 & 0.82 & 0.78 \\
     \hline
     \textbf{Value at Risk} & \textbf{-1.63\%} & -1.77\% & -1.64\% & -1.73\% & -2.13\% & -1.73\% & -1.67\% \\
     \hline
    \end{tabular}
    \bigskip
    \caption{Comprehensive statistical metrics for Sentiment-Ensemble and ensemble agents against DDPG, A2C, PPO, and TD3, as well as the DJIA.}
\end{table*}

\end{document}